# Time-Frequency Localization Using Deep Convolutional Maxout Neural Network in Persian Speech Recognition

Arash Dehghani[1], Seyyed Ali Seyyedsalehi[2*]

[1,2] Faculty of Biomedical Engineering, Amirkabir University of Technology (Tehran Polytechnic), Hafez Ave., Tehran, Iran

[1]arash.dehghani@aut.ac.ir, [2]ssalehi@aut.ac.ir

**Abstract**

In this paper, a CNN-based structure for the time-frequency localization of information is proposed for Persian speech recognition. Research has shown that the receptive fields' spectrotemporal plasticity of some neurons in mammals' primary auditory cortex and midbrain makes localization facilities improve recognition performance. Over the past few years, much work has been done to localize time-frequency information in ASR systems, using the spatial or temporal immutability properties of methods such as HMMs, TDNNs, CNNs, and LSTM-RNNs. However, most of these models have large parameter volumes and are challenging to train. For this purpose, we have presented a structure called Time-Frequency Convolutional Maxout Neural Network (TFCMNN) in which parallel time-domain and frequency-domain 1D-CMNNs are applied simultaneously and independently to the spectrogram, and then their outputs are concatenated and applied jointly to a fully connected Maxout network for classification. To improve the performance of this structure, we have used newly developed methods and models such as Dropout, maxout, and weight normalization. Two sets of experiments were designed and implemented on the FARSDAT dataset to evaluate the performance of this model compared to conventional 1D-CMNN models. According to the experimental results, the average recognition score of TFCMNN models is about 1.6% higher than the average of conventional 1D-CMNN models. In addition, the average training time of the TFCMNN models is about 17 hours lower than the average training time of traditional models. Therefore, as proven in other sources, time-frequency localization in ASR systems increases system accuracy and speeds up the training process.

# 1. Introduction

As long as the performance of the Automatic Speech Recognition (ASR) System surpasses human performance in accuracy and robustness, we should get inspired by the essential components of the Human Speech Recognition (HSR) [1]. Natural sounds have structurally rich acoustic spectra and can simultaneously vary along with spectral, temporal, and intensity, causing variations in quantities such as speaker, tone, age, accent, and environment, which have led to more research on how dynamic spectrotemporal signals are optimally processed and recognized by the auditory system of humans and animals [2]. Various biology-inspired methods have improved the ASR systems, including Perceptual Linear Prediction (PLP), Mel-scale, and spectrotemporal processing, which have received less attention from researchers. Numerous studies, including [2] [3] [4] [5] [6] [7], have been performed to identify the function of the mammalian auditory system, proving that neurons in the Inferior Colliculus (ICC) are sensitive to systematic manipulations of temporal, spectral, binaural, and intensity stimulus attributes. They were inspired by psychophysical and physiological evidence and described auditory models that combine temporal and spectral modulations [4]. Fritz et al. [6] showed that localized task-related facilitative dynamic changes in the spectrotemporal receptive fields (STRF) of mammals' primary auditory cortex in some target detection tasks enhance overall cortical responsiveness of the target tones and increase the likelihood of capturing the attended target. Based on experimental observations, Shannon et al. [8] showed that a combination of spectral and temporal cues is needed for robust speech pattern recognition as when spectral cues are disturbed, temporal cues can be used. Some works have conducted experiments using cochlear implants and showed that spectral and temporal codes in the peripheral auditory system are rich in sound-pitch information [9] [10] [11].

Research has shown that mammalian visual and auditory systems' neural structures, processes, and characteristics are similar [12] [13]. Even though most studies focus on the spatial part, visual system neurons process information in localized spatiotemporal regions [14]. According to STRF models, the spatial dimension in the visual system is functionally similar to the spectral dimension along the cochlea in the auditory system [12]. Therefore, we can use similar tools to implement these systems. It is unclear exactly how spectral and temporal acoustic dimensions are jointly processed by the brain [2]. Still, more sophisticated models for combining and localizing time-frequency information can improve the performance of ASR systems [15]. The main activities performed for time-frequency information localization of audio signals can be divided into two general parts: feature extraction and acoustic model. Time-frequency analysis has been the most critical and dominant feature representation for ASR. Much work has been done to localize an event in the time-frequency domain [16]. Feature extraction by Gabor filters, which has biological and physiological roots, has improved the performance of speech recognition systems [17]. In addition to Gabor, some methods use two-dimensional transformations such as wavelet and discrete cosine transform (2D-DCT) to localize time-frequency information [18] [19].

In the case of acoustic models, researchers have proposed many solutions for localizing time-frequency information using spatial or temporal immutability properties of tools such as Hidden Markov Models (HMMs), Time-Delay Neural Networks (TDNNs) [20], Convolutional Neural Networks (CNNs) [21], and Long-Short Term Memory Recurrent Neural Networks (LSTM-RNNs) [22]. TDNNs and HMMs were generally used to model time sequences [23]. Some works also used an augmentation of spectrogram with Delta and Double-Delta to consider the dynamics of temporal information. However, temporal information was lost because the derivation period was short [18]. CNN, an extended and optimized model of TDNNs, was introduced by Yann Lecun [24], which somehow solves the localization problem. [25] Moreover, [26] were the first works to use CNNs for audio. Many works have used CNNs in HMM/DNN hybrid models in which weight sharing was performed only along with the frequency axis [25][26][27][28][29][30], assuming that HMM can model temporal changes by its dynamic modes. As a result, weight sharing in the realm of time has not received much attention from researchers [31]. [26] Furthermore, [32] compared convolutional weight sharing in frequency and time dimensions and concluded that two-dimensional CNN (2D-CNN) in time and frequency caused a slight improvement in the recognition result. However, despite the slight improvements made by 2D-CNNs, they do not have the proper structure to localize audio information in both time and frequency domains. In the last few years, much work has been done to localize time-frequency information in the acoustic model of the ASR systems, some of which used CNNs [26] [31] [33] [34] [35] some used LSTMs [36] [37] [38] or a combination of these two structures [39] [32] [40]. In addition to ASR, time-frequency localization of audio has various applications in various tasks, including speech enhancement [41], language identification [42], acoustic scene classification [43] [44], audio super-resolution [45], audio restoration [46] and voice conversion [47].

This paper has used a structure based on CNNs to localize time-frequency sound information in the acoustic model. In this structure, two parallel blocks of 1D-CMNN networks are applied simultaneously but independently to the spectrogram, each of which has weight sharing in only one dimension. As a result, each block will perform localization in one dimension. The output of the blocks is then concatenated and applied jointly to a fully connected maxout network for classification. We have used newly developed methods and tools to improve the performance of the model. We used Rectified Linear Unit (ReLU) [48] and maxout [49] neuronal models to improve the model training process and used Dropout [50] to increase the generalization power of the model and prevent over-fitting. Weight normalization [51] was also used to prevent the model from becoming unstable during training. In the following and the second part, we will describe the background and related works. The third section presents the materials and methods used in this article. TFCMNN model will be described in Section 4. In the fifth section, we have the experiments and results, and finally, in the sixth section, we will discuss and conclude this article.

## 2. Background and related works

The motivation for the time-frequency localization of information in ASR application goes back to the experimental observations of some mammalian auditory systems. Although previous research has shown that neuronal response strength varied with the intensity and the fundamental frequency of the stimulations, it is shown in some pieces of literature such as [2] [3] [4] [5] [6] [7] that relative response to different ripple spacings remains essentially constant with changes in the intensity and the fundamental frequency [7]. Therefore, researchers have concluded that the processing in the mammalian auditory system is done so that the audio information localization is done in a spectrotemporal manner [6]. Researchers have designed models to create spectrotemporal information localization facilities in ASR systems based on these experimental results. In general, we can divide the work done concerning time-frequency localization into two parts. The first part deals with the feature extraction stage from the raw audio signal, and the second part is about the works that tried to achieve time-frequency localization using acoustic models. In the following, we describe each part in detail.

### 2.1. Feature Extraction

There are a variety of methods for feature extraction to perform time-frequency localization. Methods inspired by the human auditory system are suitable for this purpose. One of the first steps taken for the time-frequency localization of the audio signal is the use of Gabor filters. Gabor filters localize the signal in time-frequency zones that are similar to the performance of biosystems. Various works, such as [52] [12] [16] [53] [17] [54], inspired by experimental observations on the mammalian auditory system, presented a feature extraction method based on Gabor filters. In [55], first, the spectrogram is taken, and then the 2D-Gabor filters are convolved with it. The difference with CNN feature maps is that Gabor filters are fixed and not trained. Continuing the previous work, they used 1D-Gabor filters instead of 2D-Gabor filters in the fields of time and frequency [15]. The results show that time and frequency processes can operate independently and without affecting each other. They reported that converting Gabor 2D filters to 1D improved system performance in noisy conditions and reduced filters. Some works like [56] combined CNNs and

Gabor filters. In this way, Gabor filters are replaced with CNN feature maps and trained in accordance with the rest of the model. In [57], 2D-DCT and Gabor methods have been used to extract features. In addition to Gabor, other works using various methods tried to perform time-frequency localization in the feature extraction stage from the raw audio signal. In [18], a two-dimensional feature extraction method inspired by empirical research on the mammalian auditory system is presented that performs better than MFCC. In this method, STFT is taken, and then two-dimensional conversion is carried to include time-frequency composite information. Inspired by the biological system, [19] used a two-dimensional wavelet transform to extract spectrotemporal features to deal with time and frequency variations.

## 2.2. Acoustic Models

As previously mentioned, numerous studies have shown spectrotemporal localization in the receptive fields of auditory cortical neurons, which has inspired many feature extraction methods. However, recent studies show that the receptive fields of neurons in the midbrain Inferior Colliculus (ICC) also have spectrotemporal plasticity facilities improving recognition performance [58]. This finding confirms the design of acoustic models for time-frequency localization. In general, most structures designed for the acoustic model of ASR systems perform the time-frequency localization process using the spatial invariability property of CNNs or the temporal localization property of LSTM-RNNs. In most of these works, CNNs or LSTMs, or a combination of these two structures, have been used. We describe these three categories in detail below.

### 2.2.1. CNN-based Acoustic Models

One of the first works to challenge time-frequency localization in the acoustic model was [26] which compared CNN in terms of time and frequency and concluded that 2D-CNNs performed better. [59] Combined the time-domain CNN structures of [60] and the frequency-domain CNN structures of [29], proving that the hybrid model achieved better results than both of them. The method proposed in [31] has a similar approach to the method presented in this article. It uses parallel CNNs that operate in time and frequency domains. They show that parallel CNNs have improved the training process and reduced the number of filters. [35] have designed time-frequency kernels for CNN for performance stability, thereby shifting in time and frequency and the size of the kernels to fit each dimension embedded to combine time-frequency information in CNN. Kim et al. used a 3D-CNN network to meet the challenge of time-frequency dynamics localization in the application of emotion recognition [61]. These networks are invented to recognize action in video sequences [62][63]. They argued that 2D filters could not model temporal information and properties, and hybrid CNN-LSTM models have a lot of parameter volume and are challenging to train. However, 3D-CNN can improve system performance by extracting the spectrotemporal features in a sequence. In [33] and [34], 1D-CNNs have been used to derive the time-frequency property from the raw audio signal. In the first layer, each filter tries to extract spectra features. Then in the upper layers, time-domain 1D-CNNs are applied to the output of the first layer, resulting in total time-frequency immutability. They adjust the time-frequency resolutions with the dimensions and steps of the filters, and each filter has to learn a specific frequency feature. The greater the filter width over time, the more understanding about low-frequency and high-frequency bands will be almost ignored, and vice versa.

### 2.2.2. LSTM-based Acoustic Models

Inspired by the human auditory system, with this notion that RNNs can store and process sequence information, an LSTM-based network is designed [37] to operate in time and frequency. A two-stage network that operates on the frequency by F-LSTM in the first stage and the second stage operates on the time by T-LSTM. The frequency section first acts like modeling the frequency distortions and then gives the output to the T-LSTM network to stable the model over time. In [36], 2D-LSTM models the time-frequency information in the same layer and applies its outputs to a time-domain 1D-LSTM layer. To improve the performance of F-LSTMs, [38] used multi-view blocks with different steps and different window sizes and combined the output of blocks to a reduced display level. First, parallel F-LSTM with additional steps and window sizes are applied to the spectrogram, and then its output is given to T-LSTM to be localized in time. They reported that adding F-LSTM in the frequency domain to T-LSTM in the time domain has improved the performance of the ASR system.

### 2.2.3. Hybrid CNN-LSTM-based Acoustic Models

In LSTM-CNN hybrid models, LSTMs have been used to model temporal information in many works. This way, after applying CNNs over the frequency axis, its output is applied to LSTM networks over the time axis. However, these models have a lot of parameter volume, and their training is difficult [61]. 2D-CNN and LSTM have been used in (Amodei et al., 2016) and slightly improved the recognition result. It has been concluded that the first layer should have CNN on the frequency axis because upper LSTM-RNN layers would eliminate frequency. [39] used CNNs for frequency modeling and 1D-LSTMs for temporal modeling in a hybrid network with DNNs. In this structure, LSTM models temporal variations and spectral variations are modeled by CNN, and their result is applied to a fully connected network to be classified. [40] use 2D-RNN and CNN in the frequency domain in the TDNN structure to reduce input variations in the time and frequency domains.

## 3. Material and Methods

In this section, we describe the material and methods used in this article.

### 3.1. Convolutional Neural Networks (CNN)

The most crucial disadvantage of fully connected neural networks is that they do not have a mechanism to deal with variations and distortion in input data. Image characters, speech signal spectra, and other one- or two-dimensional signals must be approximate in size and concentrated in the input space before being sent to the first layer of a neural network. Unfortunately, no such premise can be complete. Words can be spoken at different speeds, steps, and accents, causing differences in distinctive features in the input data. Another disadvantage of fully connected structures is that the input topology is wholly ignored. Input variables can be displayed in any order without being affected by the training outcome. Whereas the image or spectrum representing speech has a solid two-dimensional local structure, the time-series signals have a one-dimensional structure. The variables or components of signals, which are spatial or temporal, are also very closely related. Local dependence is the reason for the advances made in extracting and combining local features before considering the spatial or temporal nature of the data. Convolutional Neural Networks (CNNs), presented in 1995 by Yann Lecun [24], extract local features by restricting the input field of neurons, forcing them to be local. In other words, in CNNs, spatial immutability will be realized automatically by the forced repetition of weight configurations in space. We can consider CNN, which has weight sharing over time, as a broad version of TDNN. Yann Lecun evaluated their performance in image and audio processing applications and obtained good results from them. Since then, this structure has had a high ability to achieve immutability by spatial transfer and localization of patterns in the category of image processing [29] and speech processing [25] [26] and achieved excellent results.

CNNs combine three structural principles to achieve spatial immutability: Local receivers, Shared weights, and spatial and temporal sampling and integration. Each layer receives inputs from a group of neurons in the previous layer located in small, contiguous locations. The idea of connecting units to local parts in the perceptron input space dates back to Hubel and Wiesel's work [64], in which they made discoveries about the functional architecture of the cat's visual system. With locally received areas, neurons can detect and extract basic visual features such as oriented edges, corners, endpoints, or local features in the speech spectrum. These features will finally be combined in the upper layers. The weight sharing method is the main factor in reducing the number of free parameters of the network, reducing the system's volume, and improving the network's performance [24]. There are two weight-sharing approaches in CNNs that create two types of structures: One-Dimensional Convolutional Neural Networks (1D-CNN) and Two-Dimensional Convolutional Neural Networks (2D-CNN). Figure 1 shows a sketch of these two structures. The term two-dimensional means that sharing weights in the convolutional layer takes place along two dimensions. In other words, the receptive field of neurons in each map can transmit to both sides. Nevertheless, the term one-dimensional means that weight sharing is done only along one axis so that the receiving field of neurons in feature maps is transmitted only along one axis. As a result, feature maps only expand in one direction. When applied to any dimension, the weight-sharing process can make the model flexible against slight spatial variations in that dimension, resulting in system stability against irregularities along that dimension.

### 3.2. Pre-training

Due to many local minima, DNNs will usually not converge [65]. However, with proper initialization of network weights, many local minima can be avoided. Pre-training methods are used to find the initial values of network weights and free the learning process from the local minimums in the middle of the road as a fundamental obstacle in the training process. These methods seek to find an appropriate starting point for network weights and, in addition to facilitating the network training process, also improve the generalizability of the network [66]. In 2006, Hinton proposed the Restrict Boltzmann Machine (RBM) method for pre-training multilayer neural networks to reduce the non-linear dimension [67]. In this method, the multilayer network is broken down to the corresponding number of RBMs, and the pre-training of the weights is done through these RBMs. In 2015, Seyyed Salehi et al. introduced the layer-by-layer pre-training method for pre-training Deep Bottleneck Neural Networks (DBNNs) to extract the principal components [65]. However, we used a bidirectional version of this method to pre-train DNNs [66]. This method is used to converge fully connected networks with neurons with sigmoid and sigmoid tangent nonlinearity. However, using these methods will not significantly improve the model's performance with the advent of more efficient neuronal models.

## 3.3. Neuronal models
### 3.3.1. Rectified Linear Unit (ReLU)

Based on the biological model of neurons presented by Dayan and Abbott in 2001 [68], Glorot et al. showed that using ReLU neuronal model in Artificial Neural Networks (ANNs) instead of hyperbolic tangent neuronal models would improve their performance [48]. Despite being non-linear hard and non-derivative at point zero, it is more biologically similar to natural neurons and enhances the function of ANNs and their training process. Its approximate equation is as follows:

$$h^{(i)} = \max(w^{(i)T}x, 0) = \begin{cases} w^{(i)T}x &, w^{(i)T}x > 0 \\ 0 &, \text{else} \end{cases} \quad (1)$$

ReLU neuronal model, like biological neurons, creates sparsity in the network. Due to the saturation at zero, the training process may seem to be disrupted, which is why an extended model of this neuron was presented called Soft-Plus [48], which has a softer nonlinearity than the original model. Ziglar et al. in 2013 used this neuronal model for speech recognition and obtained good results [69]. The ReLU neuronal model has been used in many speech recognition applications and has performed better than previous structures [70]–[73]. This neuronal model has better performance without using biases [69]. Also, due to the instability created by its linear part in the network, weight normalization and sometimes layer normalization have been used [72].

### 3.3.2. Maxout

As mentioned earlier, the ReLU neuronal model suffers severely from zero saturation and divergence in its linear region. Although improved structures such as Soft-Plus could cope with these problems to some extent, such models did not eliminate these problems. They were continuously subject to saturation and divergence. In 2013, Goodfellow et al. introduced a model called maxout [49]. Despite its simplicity, this model essentially eliminated the shortcomings of ReLU. Its name is since its output is a maximum of a group of neurons and is somehow accompanied by Dropout [50]. By removing the saturation of ReLU neurons, maxout creates the ground for better network training and easier convergence. The maximization process is also considered a feature selector. The advantage of these neurons is that, unlike ReLU and sigmoid neurons, they always pass through the gradient and do not cause it to degenerate. This property is because its output at any time is equal to the output of a neuron with a linear function that has a maximum value relative to a group of neurons, so the derivatives will always be equal to one. The maxout model is simply a feed-forward structure, such as a multilayer perceptron or deep CNN that uses a new operator function that maximizes feature maps' output. This neuronal model will implement with the following operations:

$$h_i(x) = \max(z_{ij}) \quad i \in [1, k] \tag{2}$$

$$z_{ij} = x^T W_{\ldots ij} + b_{ij}, \text{and } W \in R^{d \times m \times k}, \text{and } b \in R^{m \times k} \tag{3}$$

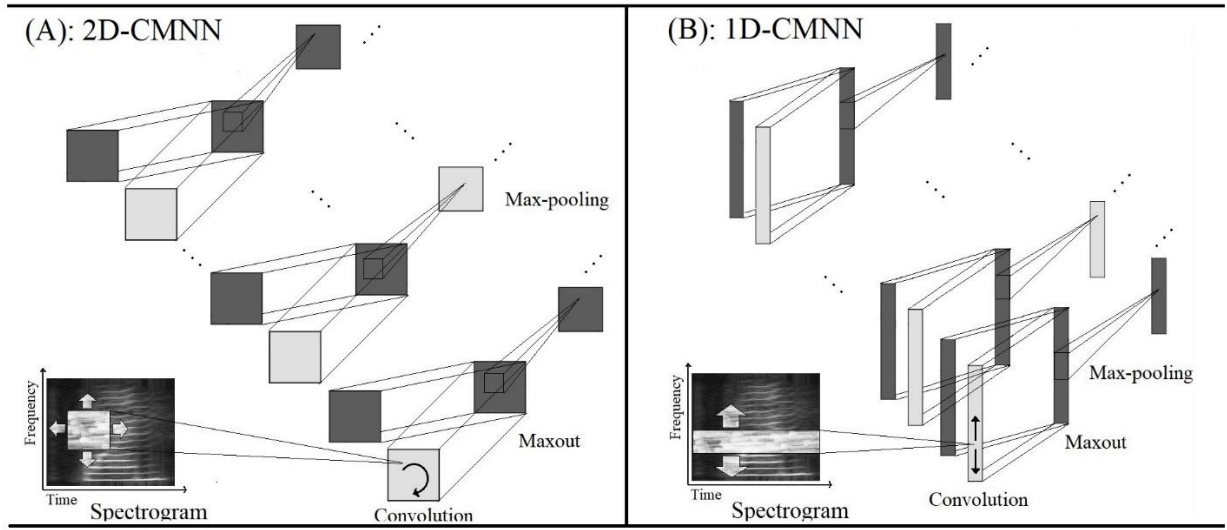

*Figure 1. Maxout structure in 1D and 2D-CMNN. (A) shows the first layer of a 2D-CMNN with maxout neurons, and (B) demonstrates the first layer of a 1D-CMNN with maxout neurons. In this view, the Maxout box contains two feature maps, from which it selects the maximum for each element located in the feature map. As we can see in the figure, weight sharing in one-dimensional structures is done in one dimension only, and CNN filters are shifted along one dimension on the spectrogram, although in two-dimensional structures, they are shifted in both dimensions.*

Where $h$ denotes the output of the maxout unit and $x$ represents the inputs. As shown in Figure 1, a maxout feature mapping can be constructed by maximizing the k independent feature maps in a CNN. When instructed by the Dropout method, we multiply the input elements to the Dropout mask before they are multiplied by the weights and reach the maxout operator. A maxout unit can be interpreted as a piecewise linear and approximate model of an arbitrary convex operator. Maxout neurons learn the relationship between hidden units and the function of each hidden unit. Maxout grid with $k$ hidden units can approximate any continuous function, of course, when k tends to infinity. Therefore, it provides the basis for many conventional operators in terms of design. In other words, maxout neurons can simulate different functions. In general, the output of these neurons does not have sparsity. However, the gradient is remarkably sparse, and the Dropout will artificially sparse its effective display during training. This neuronal model is specifically designed to facilitate Dropout optimization operations and improve the fast Dropout averaging model [49]. Various research groups used the maxout model in the structure of their acoustic model for ASR [74][75][76]–[78] and got better results than the previous structures.

### 3.4. Regulators
#### 3.4.1. Dropout

Deep neural networks (DNNs) with non-linear functions at different layers can learn a complex mapping between inputs and outputs. However, with relatively limited data volumes, learning this complex relationship between finite data sets can become problematic and impair the network's ability to generalize to unseen data sets. As a result, these complex mappings between trained data will not be generalizable to test data. This phenomenon will lead to the problem of over-fitting [50]. Many methods have been proposed to solve this problem, and it can be said that the oldest and most important of them is the Bagging method [79]. In this method, various neural networks are trained on a data set, and their results are averaged during testing. However, there is another way to have fewer calculations, share information between training models, and predict the results with a more efficient averaging method. Dropout, presented by Srivastava et al. in 2013 [50], is a method that provides a solution to all these problems. This method offers a way to combine different models with different structures with a more efficient averaging process, which prevents over-fitting. The name Dropout is derived from randomly removing neurons from the network structure during training. Each neuron with a probability $P$ will be present in the network. A sparse network will be obtained after removal. In other words, we train a thinner network instead of the main network in each iteration. When testing, those neurons that were present in the network with a probability of $P$, are multiplied by $P$. As a result, a neuron that is more likely to be present in the network during training has a more significant impact on the network and should also have a greater effect on the main network during testing. To implement this method, consider a DNN with $L$ layers in which $l$ is assumed to be a member of the set $[1, 2, 3, ..., L]$ specifies the number of hidden layers. For a standard feed-forward network, the general form of network equations for this method is as follows (4-6):

$$z_i^{(l+1)} = w_i^{(l+1)} y^l + b_i^{(l+1)} \tag{4}$$

$$y_i^{(l+1)} = f\left(z_i^{(l+1)}\right) \tag{5}$$

$$y^{(l)} = r^{(l)} * y^{(l)}, \text{ and } r_j^{(l)} \sim Bernoulli(p) \tag{6}$$

Where $f$ is an arbitrary function. The vectors $z_i$, $y_i$, $b_i$, and $w_i$ represent the input, output, weight, and bias vectors for layer $l$th, respectively. At the time of testing, the weights of the $l$th layer are also calculated through equation (7), and no neuronal removal is performed.

$$W_{test}^{(l)} = pW^{(l)} \tag{7}$$

### 3.4.2. Weight Normalization

As long as we use limited-function operators such as sigmoid functions, the neurons' output and weights are always bounded and will not reach infinity. However, when we use neural models that do not have a limited output function range and can generate huge numbers, the risk of instability towards infinity will threaten the network at any time. To protect the network against instability so as not to disrupt the training process, we must limit the vector of the weights and output of neurons to keep their directions unchanged. The Weight Normalization method is suitable for this purpose [51]. In this method, the size of the network weight vector is limited to a fixed number such as $C$ and does not allow this number to expand. According to equation (8), we enclosed the magnitude of the weight vector in a hyperdimensional sphere with a radius of $C$, which is the maximum rate.

$$\|W\| = [\textstyle\sum_i |e_i|^2]^{1/2}, \ \|W\| < C \tag{8}$$

In (8), $i$ specifies the number of elements of the vector $W$, and $e$ denotes the numerical value of each component. As long as the size of the weight $W$ is less than constant $C$, no action will be taken on the weight vector. Nevertheless, when the size of $W$ rises from $C$, the weight vector values are corrected so that its magnitude will be equal to $C$ without any change in direction. The advantage of this method is that we can increase the learning rate without fear of excessive weight gain, lack of convergence, and network instability. This feature allows us to start training with a greater learning rate, access more weight space, and smooth out previously difficult areas with Dropout's noise [50].

## 4. Proposed Structure (TFCMNN)

Various methods have been proposed for the time-frequency localization of speech signal information to improve the performance of speech recognition acoustic models. In the previous section, some of these methods were described which in most cases, CNNs and LSTM-RNNs were used. The most important reason for using these structures is the network's strengthening against minor variations along the speech spectrum's time and frequency dimensions. LSTMs have a great ability for modeling time series, and using them in the acoustic model of speech recognition system will improve performance. However, they alone cannot model the time-frequency dimensionality of speech signals and will perform better if combined with CNNs in the lower layers of the network. Nevertheless, as mentioned in other sources, a hybrid network has a huge parameter volume and many calculations compared to the rate of performance improvement. Therefore, despite the partial improvement of system performance, their use has high costs. Time-frequency localization in ASR systems has improved system performance, reduced the number of parameters, and reduced calculations and training time. Some works proved that extracting features from the speech signal in separate phases would improve the time-frequency localization [15] [31]. This work uses a structure based on 1D-CMNN to jointly combine and localize local small-signal variations at the acoustic model's time and frequency dimensions. The proposed model's overall design is shown in Figure 2. Our previous work found that maxout neurons have a higher generalization power used in this structure [80].

In this structure, Time-Frequency Convolutional Maxout Neural Network (TFCMNN), time-domain and frequency-domain 1D-CMNN blocks operating in parallel were trained simultaneously in conjunction with the fully connected maxout network, one of which shares weights along the time and the other shares weights along with the frequency. As shown in Figure 2, the upper block operates in the time domain, and the lower block operates in the frequency domain. Finally, extracted features of the upper layers of these two networks are concatenated and applied jointly to a fully connected maxout network for the classification process. All weights were trained by the error back-propagation algorithm in which the error signal is transmitted from the fully connected maxout network to the parallel 1D-CMNN blocks.

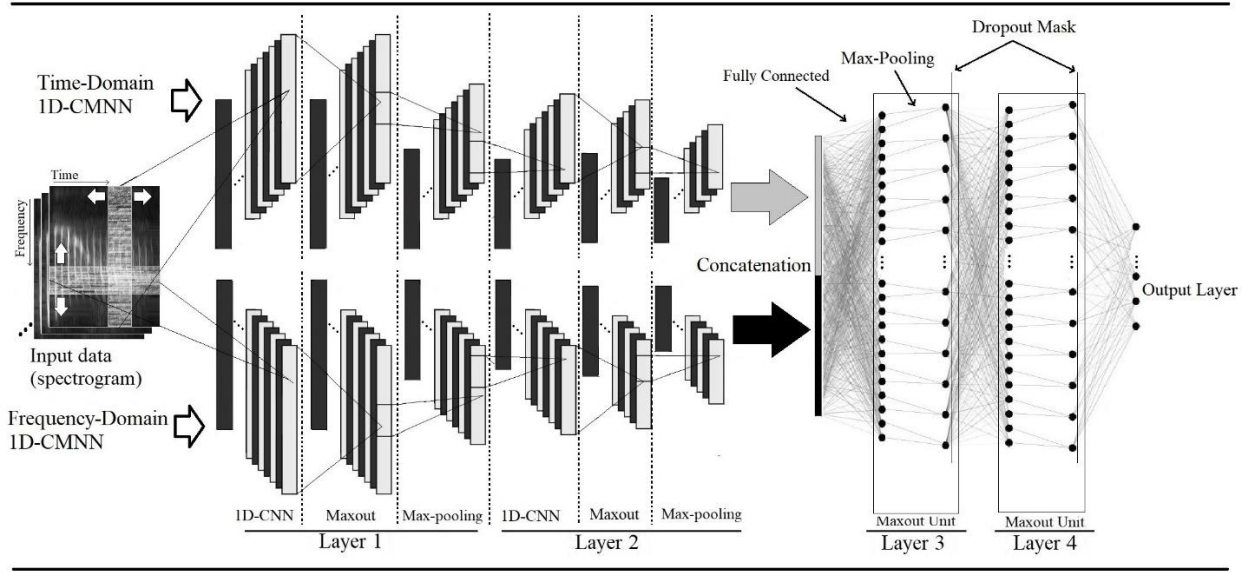

*Figure 2. Demonstration of the proposed TFCMNN structure. The details of each layer are distinguished in the figure. Some of the maxout layers' dashed lines have been removed to avoid image clutter.*

In this structure, two layers of CMNN, including a 1D-CNN layer, a maxout layer, a max-pooling layer, and two fully connected maxout layers, are used. We use Dropout only for fully connected maxout layers. The parallel CMNNs separately model variations and displacements in time and frequency, and somehow the network is resilient in both dimensions. Compared to other models, the advantages of this structure include increased recognition accuracy and a decrease in the computational volume and training time.

## 5. Experiments and Results

This section will first introduce the FARSDAT Persian speech dataset and then describe the experiments performed and the settings applied in more detail.

### 5.1. FARSDAT Speech Dataset

FARSDAT [81] is a speech database of Farsi spoken language which contains continuous and clear Persian speech signals from 304 male and female speakers who differ in age, accent, and level of education. Each speaker read 20 sentences in two parts. The speech was sampled at 44.1 kHz by 16-bit sound Blaster hardware on IBM microcomputers. These data are fragmented and labeled at the phoneme level with 23-millisecond windows, and with the progress step of half the length of these windows are stored as separate files. Labeling of FARSDAT databases has been done by people familiar with linguistics and with the help of relevant software. These data are internationally recognized as standard Persian language speech data and train intelligent speech recognition devices. In all the experiments performed in this work, 297 speakers randomly selected from 304 people are considered train data, and the speech of the remaining seven speakers is used as test or evaluation (Eval) data. Development (Dev) data are randomly selected from train data.

We use preprocessors to extract the feature from the raw signal to remove additional information from the speech signal and obtain the most necessary information needed for separation and classification [27]. Various methods such as Perceptual Linear Prediction Coefficients (PLP), Mel-Frequency Cepstral Coefficients (MFCC), and Logarithm of square Hanning Critical Filter Bank Coefficients (LHCB) are present for feature extraction. According to the results reported by Rahiminejad in 2003 [82], feature extraction from FARSDAT data using LHCB parameters will be better than other methods, so we used this method to extract features from raw speech signals. The LHCB method is one of the spectral methods for extracting Bark-based representation parameters. The characteristic vector of each frame is obtained using a Hanning Critical Filter Bank. After sampling the speech signal and eliminating the DC values of the frame, it is multiplied by the Henning time window, and then, a short time Fourier transform will be taken. After calculating the power spectrum in the next step, Hanning Critical Filter Bank will be applied to the power spectrum. Finally, the logarithm of the output of each filter is taken. A total of 18 representation parameters will be extracted for each frame. The obtained parameters will reduce the volume of speech signal information and prevent additional information.

Nevertheless, to train neural networks with these parameters, each of the 18 representation vectors must be normalized. Lack of normalization will make the model training process more difficult. There are many normalization methods. According to Rahiminejad [82], Norm-2 normalization with a variance of 0.5 has the best performance among other normalization methods, so we used the same method for normalization. Each frame of the speech signal spectrogram, consisting of 18 parameters, has its phonetic label, which specifies which phoneme represents the 29 phonemes of the Persian language and silence. However, it is common to train the model by a few frames before and after the mainframe. For this reason, the spectrogram must be windowed. In most of the experiments performed in this work, the window lengths were 15 and 18, and in some cases, 12.

## 5.2. Experiments, Settings, and Results

All implementations have been done using the MATLAB program. We used the toolbox published by Palm [83] as the central core of implementations. The structures not available in this toolbox, such as 2D- and 1D-CMNN, maxout networks, Dropout, and weight normalization, were added to the toolbox. We transformed 2D-CNN into 1D-CNN, added maxout neurons, and replaced the Max-Pooling with the Mean-Pulling layer. Also, due to the inadequacy of the programming code for the desired tasks, we optimized the code and the program implementation process. Considering the previous work [80], the application of the Dropout method in CNN layers had little effect, so Dropout masks were applied only to the output of fully connected maxout layers. Also, the number of neuronal units in the maxout box (maxout pieces) is considered 2 in most structures, but in some cases, it was 3. The experiments are divided into two categories to measure the performance of the TFCMNN model compared with conventional 1D-CMNN models. The first category of experiments was performed on conventional 1D-CMNN structures, and the second was performed on TFCMNN structures. In the first category of experiments, we used conventional 1D-CMNNs for localization in a single dimension.

*Table 1- Experimental results of the first category of experiments: The best selected conventional 1D-CMNNs structures with weight sharing in time or frequency dimensions on the FARSDAT speech dataset.*

| Ex. ID | Structure | Weight sharing | Epoch | Training Time | Frame Accuracy (Dev) | Frame Accuracy (Eval) | Phoneme Error Rate (%PER) |
|---|---|---|---|---|---|---|---|
| A-1 | C40 K5 S2 C60 K4 S2 F400 F400 | Frequency | 10 | 84 h | 90.03% | 85.37% | 17.93 |
| A-2 | C40 K7 S2 C40 K3 S2 F600 | Frequency | 20 | 77 h | 91.85% | 86.52% | 16.78 |
| A-3 | C40 K5 S2 C40 K4 S2 F400 | Frequency | 14 | 60 h | 91.56% | 86.30% | 17.03 |
| A-4 | C40 K3 S2 C40 K3 S2 F600 | Frequency | 14 | 65 h | 92.16% | 86.61% | 16.69 |
| A-5 | C40 K7 S2 F400 F400 | Frequency | 20 | 46 h | 93.17% | 87.89% | 15.41 |
| A-6 | C40 K7 S2 F400 F400 | Frequency | 16 | 74 h | 92.83% | 87.26% | 16.04 |
| A-7 | C100 K3 S2 F400 F400 | Time | 16 | 66 h | 92.37% | 87.97% | 15.33 |
| A-8 | C100 K7 S2 F400 F400 | Time | 14 | 59 h | 93.14% | 87.98% | 15.32 |
| A-9 | C40 K3 S2 C40 K3 S2 F600 | Time | 16 | 65 h | 93.45% | 88.08% | 15.22 |
| **Average** | | - | 14.9 | 73.6 h | 92.18% | 86.89% | **16.41** |

In general, the initial settings and parameters of the models are adjusted based on the results of our previous work [80]. In order to achieve the most optimal settings of the models, many experiments were performed. By comparing the obtained results, the optimal parameters of the models, such as window sizes, number of neurons in the maxout box, and the probability of the presence of neurons in the Dropout method, were obtained. To obtain the best structures of 1D-CMNNs, as well as to evaluate and compare the performance of structures that have weight sharing along with time or frequency, many experiments were performed on structures with variations in the number of layers, number of neurons, number of feature maps, maxout box units and different window sizes. Finally, the best structures were obtained from all these experiments. Table 1 shows the results of the optimal structures. In all experiments, the learning rate was initially assumed to be 0.1, and Max-Norm weight normalization was efficient at 0.8. The batch size was chosen to be 100 in most cases. In each epoch, we evaluated the performance of the model on Eval. data. When the recognition score was less than that obtained from the previous epoch, the learning rate was divided by 2. If this happened five times, the network training process would be stopped automatically, and the results would be saved. In Table 1, the structure of the models, the number of convergence epochs, the training time, and the evaluation metrics on the Eval. and Dev. data based to equation 9 and equation 10 and the dimension along which weight sharing is performed are considered as comparative quantities. We used two evaluation metrics to evaluate the performance of the models during training and testing, including the frame recognition criterion and Phoneme Error Rate (PER). The frame recognition criterion is calculated as follows:

$$\text{Frame Accuracy} = (\frac{c}{N}) * 100 \tag{9}$$

Where $c$ is the number of correctly detected frames by the model and $N$ is the number of total frames in each utterance. Also, the Phoneme Error Rate (PER) criterion was used, which is calculated based on the following equation:

$$PER = (\frac{S-D-I}{N}) * 100 \tag{10}$$

where $N$ is the total number of labels in the reference utterance and $S$, $D$, and $I$ are substitution, deletion, and insertion errors. At the end of the table, we compute the average performance of all structures for each parameter compared with the average statistics of the second set of experiments.

Table 2- Experimental results of the second category of experiments: The best selected TFCMNN structures on the FARSDAT speech dataset.

| Ex. ID | Structure | Dropout | Epoch | Training Time | Frame Accuracy (Dev) | Frame Accuracy (Eval) | Phoneme Error Rate (%PER) |
|---|---|---|---|---|---|---|---|
| B-1 | C40 K3 S2 F400 F400 | - | 10 | 61 h | 94.58% | 87.95% | 15.35 |
| B-2 | C40 K7 S2 F400 F400 | D = 0.3 | 15 | 66 h | 94.68% | 87.97% | 15.33 |
| B-3 | C40 K5 S2 F400 F400 | - | 10 | 63 h | 94.78% | 88.25% | 15.05 |
| B-4 | C80 K7 S2 F400 F400 | - | 8 | 70 h | 94.27% | 88.57% | 14.73 |

| Ex. ID | Structure | Dropout | Epoch | Training Time | Frame Accuracy (Dev) | Frame Accuracy (Eval) | Phoneme Error Rate (%PER) |
|---|---|---|---|---|---|---|---|
| B-5 | C60 K7 S2 F400 F400 | D = 0.5 | 12 | 51 h | 94.65% | 88.58% | 14.72 |
| B-6 | C40 K7 S2 F400 F400 | D = 0.5 | 12 | 31 h | 95.38% | 88.88% | 14.42 |
| B-7 | C40 K7 S2 F400 F400 | D = 0.7 | 12 | 53 h | 95.67% | 89.42% | 13.88 |
| Average | | - | 11.2 | 56.42 h | 94.85% | 88.51% | 14.79 |

The structure of the networks is briefly stated. In this acronym, *C* indicates the CNN layer, *K* denotes the filter width for feature maps, *S* indicates the size of the pooling window in the max-pooling layer, and *F* shows fully connected maxout layers in the network. The numbers next to the characters indicate their quantity. For example, *C*40 means 40 feature maps in a CNN layer. We selected the model quantities according to the volume of training data. The number of feature maps in all CNN layers was selected from 40 to 100, and the number of neurons in all fully connected maxout layers was selected from 100 to 800. The filter widths for time-domain and frequency-domain CMNN models were selected from 3 to 7. The second category of experiments was performed on the TFCMNN structures with the same settings and training strategy we have used in the first category. Various experiments with different model quantities were performed on the FARSDAT speech dataset to compare the most optimal models. Table 2 shows the results of these implementations. To optimally display the results, we used the data stored by the models during training and displayed them graphically in Figure 3 using MATLAB and Paint Editor programs. In this figure, the vertical axis represents the percentage of frame recognition, and the horizontal axis specifies the number of epochs. To not clutter the figure, only the results of the top three models of each structure are displayed. The models related to the proposed structure (TFCMNN) are marked in green, and the models related to the conventional structure (1D-CMNN) are marked in red.

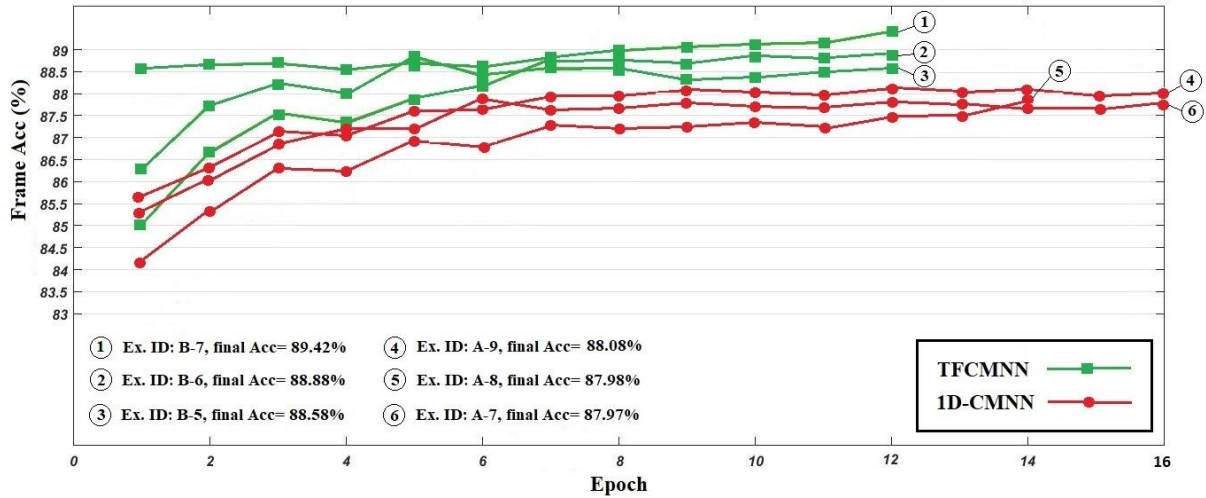

*Figure 3- Frame recognition percentage chart of the top three models of each structure in each epoch during training.*

## 6. Discussion and Conclusion

Natural sounds have rich spectral and temporal acoustic sources and can vary simultaneously in the dimensions of frequency, time, and intensity. Inspired by the biological auditory system, ASR systems deal with these distortions by increasing the likelihood of capturing events in the receiving fields and acoustic models, known as time-frequency localization. However, how spectrotemporal information is localized in biological systems is still unclear. Various structures have been proposed for time-frequency localization in ASR systems. This paper proposed the TFCMNN structure that embeds the time-frequency localization facilities in the acoustic model. The presented structure is based on CNNs and consists of two parallel time-domain and frequency-domain 1D-CMNN and a fully connected maxout network. According to the TFCMNN structure, the variations and displacements in the time and frequency dimensions will be localized separately by the parallel 1D-CMNN blocks, and the model will be resistant in both dimensions. Methods and tools such as Dropout, maxout, and weight normalization were used to improve the model's performance. We have designed two sets of experiments to evaluate the performance of this structure concerning conventional 1D-CMNN structures. All experiments have been performed with the same settings and procedures on the FARSDAT Persian speech dataset. According to the results reported in Table 1 and Table 2, the average recognition score of TFCMNN models on Eval. data is about 1.6% higher than the average of conventional 1D-CMNN structures. Also, the average training time and the number of convergence epochs for TFCMNN models are about 17.18 hours and 3.7 epochs less than conventional 1D-CMNN models. Therefore, as stated in other sources, we can say that the TFCMNN structure increased the system's accuracy and caused faster convergence.

# Acknowledgment

The authors of this article express their gratitude to Ms. Soraya Rahimi, Ms. Fatemeh Maghsoud Lou, and Mr. Arya Aftab for their valuable contributions to implementing and analyzing the performance of CNNs, maxout networks, and dropout training method. This research did not receive any specific grant from funding agencies in the public, commercial, or not-for-profit sectors.